\documentclass[reprint,superscriptaddress,aps,prb,twocolumn,floatfix]{revtex4-2}
\usepackage[utf8]{inputenc}
\usepackage{textcomp} 
\usepackage{setspace}
\usepackage{amsmath}
\usepackage{breqn}
\usepackage{graphicx}
\usepackage{verbatim}
\usepackage{float}

\usepackage{amsfonts}
\usepackage{amssymb}
\usepackage{xcolor}
\usepackage{soul}
\usepackage{tabularx}
\setcitestyle{super}
\usepackage[colorlinks,linkcolor=blue,anchorcolor=blue,citecolor=blue,urlcolor=black]{hyperref}
\DeclareGraphicsExtensions{.pdf,.eps,.png,.jpg,.mps}
\usepackage{booktabs}
\usepackage{makecell}
\usepackage{threeparttable}

\begin{document}

\title{Self-sustained microcomb lasing in an integrated hybrid oscillator}
\author{ Bitao Shen$^{1,\dagger}$, Huajin Chang$^{1,\dagger}$, Junhao Han$^{1,\dagger}$, Yimeng Wang$^{1}$, Xuguang Zhang$^{1}$, Haoyu Wang$^{2}$, Zihan Tao$^{1}$, Ruixuan Chen$^{1}$, Yandong He$^{2}$, Haowen Shu$^{1,*}$, and Xingjun Wang$^{1,3,4,*}$ \\
\vspace{3pt}
$^1$State Key Laboratory of Photonics and Communications, School of Electronics, Peking University, Beijing 100871, China.\\
$^2$School of Integrated Circuits, Peking University, Beijing 100871, China.\\
$^3$Peking University Yangtze Delta Institute of Optoelectronics, Nantong 226010, China.\\
$^4$Frontiers Science Center for Nano-optoelectronics, Peking University, Beijing 100871, China.\\
$^\dagger$These authors contributed equally to this work \\
\vspace{3pt}
Corresponding authors: $^*$haowenshu@pku.edu.cn, $^*$xjwang@pku.edu.cn.}



\date{\today}

\maketitle
\noindent
\textbf{Abstract} \\
\textbf{Microcavity optical frequency combs (microcombs) are compact, coherent light sources whose chip-scale integrability is poised to drive advances in metrology, communications, and sensing. Among available microcomb generation methods, hybrid cavities uniquely co-locate gain and Kerr dynamics, where the lasing mode directly resonates in the nonlinear microcavity, simultaneously enabling self-sustained and highly efficient microcomb generation. However, their implementation is often limited by partial integration or the need for external injection, which complicates operation architecture, raises power and hampers system miniaturization. In this work, we present a fully integrated hybrid cavity for self-sustained microcomb generation, relying solely on the co-oscillation of lasing and Kerr nonlinearity without external driving. The system collapses the pump laser, nonlinear resonator and feedback loops into a minimalist on-chip two-element cavity, consisting of a high-Q microresonator with engineered intracavity reflection and a reflective semiconductor optical amplifier (RSOA). The scheme delivers self-starting operation and stable performance without active feedback. The generated coherent microcomb achieves intrinsic linewidths below 1 kHz and integrated linewidths around 100 kHz, with self-sustained operation exceeding 24 hours. This ultra-compact architecture provides a practical path toward scalable, coherent multi-wavelength sources for integrated photonic systems.}

\vspace{3pt}
\noindent
\textbf{Introduction} \\
\noindent Microcavity optical frequency combs \cite{gaeta2019photonic, kippenberg2011microresonator}, or microcombs have seen significant development over the past two decades, due to their ability to generate broadband, high-coherence optical frequency combs in a compact form \cite{xiang2021laser, herr2014temporal}. Building on the foundations established by conventional mode-locked lasers \cite{fortier201920} and enabled by advances in integrated photonics \cite{armani2003ultra,liu2021high, xiang20233d}, microcombs, particularly integrated microcombs, have been employed in a range of applications \cite{obrzud2019microphotonic, suh2016microresonator, trocha2018ultrafast, xu202111}, including optical clocks \cite{newman2019architecture}, microwave photonics \cite{sun2024integrated, wu2018rf}, high-capacity optical communications \cite{shu2022microcomb, riemensberger2020massively} and interactions between photon and other particles \cite{yang2024free}.

Since the beginning, efforts to simplify microcomb stimulation \cite{guo2017universal} have never ceased for practical applications. In principle, microcomb can be sustained through the balance between dispersion and Kerr nonlinearity, as well as between the dissipation and gain under an external continuous wave (CW) pumping \cite{kippenberg2011microresonator, del2007optical}. However, achieving a mode-locked microcomb in practice remains a significant challenge \cite{li2017stably, liu2014investigation, herr2012universal}. For example, complex tuning or pumping strategies \cite{stone2018thermal, yi2016active, zhang2019sub, zhou2019soliton} are often required to overcome intracavity thermal effects in an anomalous dispersion cavity. Apart from Kerr nonlinearity, various interaction processes, including Raman scattering \cite{li2024ultrashort}, Brillouin scattering \cite{jia2020photonic, zhang2024strong, bunel2025brillouin} and quadratic nonlinearity \cite{bruch2021pockels, stokowski2024integrated}, have been explored for microcomb generation, with improved pulse duration, coherence or conversion efficiency. Notably, Brillouin scattering provides simple and stable access \cite{bai2021brillouin} to the single soliton state by directly positioning the Brillouin laser at red detuning where the soliton state exists. Despite these advances, these strategies still rely on the challenging fine-tuning process \cite{opavcak2024nozaki}, where an external narrow-linewidth laser must be precisely detuned relative to resonance. A simpler microcomb stimulation method is needed, where the microcomb can self-start and self-sustain, without the need of complex tuning processes and expensive external sources.

\begin{figure*}[ht]
\centering
\includegraphics[width = 17.3cm]{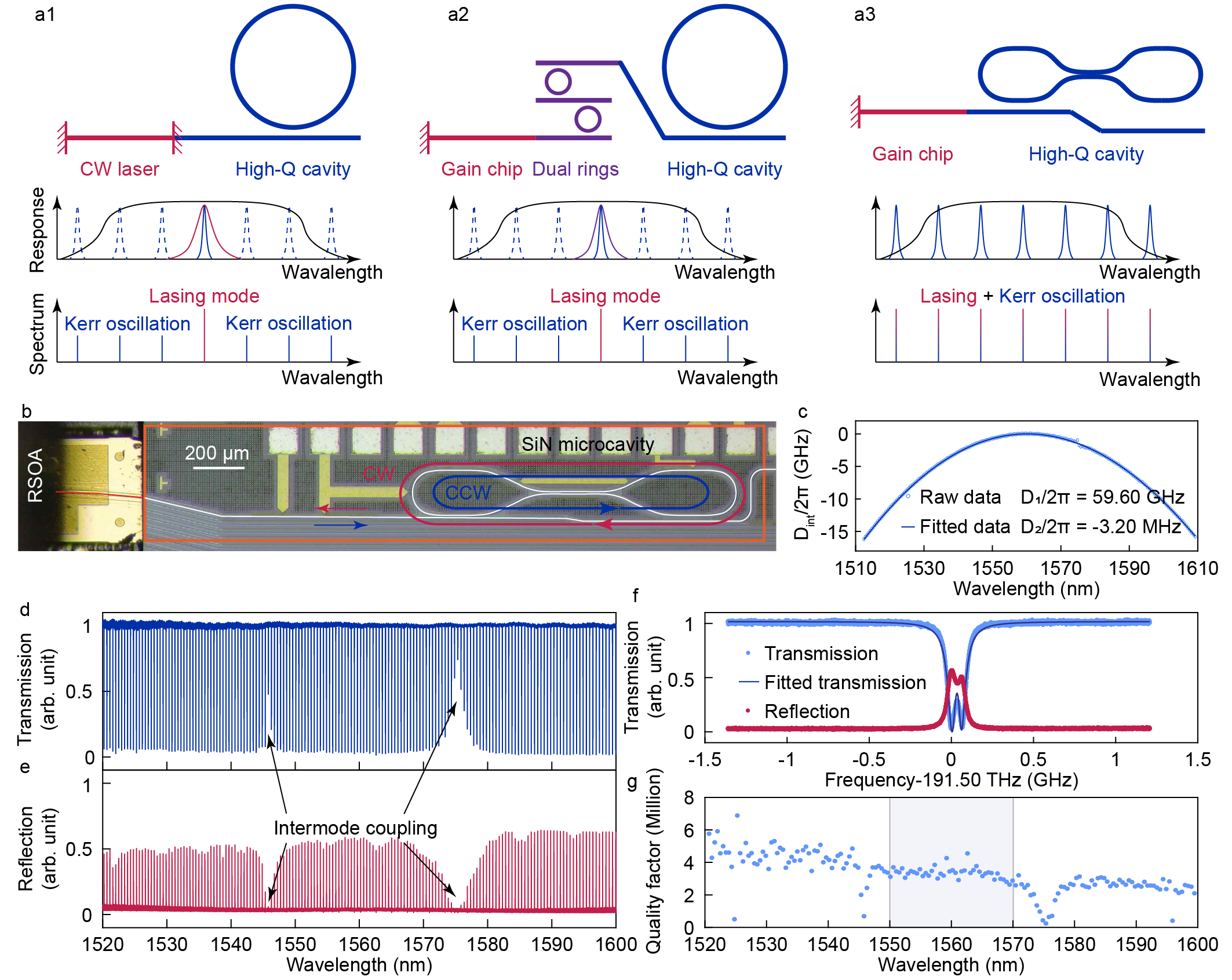}
\caption{\textbf{Principle and basic characterization. a,} Comparison of different hybrid cavity schemes of self-injection locking \cite{shen2020integrated, gil2025high} (a1), dual ring filtering \cite{stern2018battery} (a2) and the one in this manuscript (a3). Top: device structure. Middle: optical responses of different devices, with he black line indicating the gain spectrum. The red, purple and blue curve represent the filter response of the CW laser cavity in a1, dual rings in a2 and the high-Q cavity respectively. Solid and dash lines indicates whether lasing modes can be built up. Bottom: generated spectra from different schemes. \textbf{b,} Photo of a hybrid oscillator. The white and red lines show the photonic circuits on the SiN chip and the RSOA chip respectively. The blue and pink lines denote the propagation directions of the CCW and CW fields, respectively. \textbf{c,} The normalized transmission spectrum of the microresonator. \textbf{d,} The normalized reflection spectrum of the microresonator. \textbf{e,} Integrated dispersion profile. \textbf{f,} The transmission and reflection spectrum of the resonance around 1566.5 nm (191.50 THz). \textbf{g,} Loaded quality factor of resonances.
}
\label{fig1}
\end{figure*}

Hybrid oscillation processes, combining lasing and Kerr nonlinearity \cite{kondratiev2020numerical, bao2019laser,johnson2014microresonator, opavcak2024nozaki}, show promise for self-starting and sustainable microcomb operation. One of the most well-developed strategy is self-injection locking\cite{pavlov2018narrow, dmitriev2022hybrid}, where a laser cavity is coupled to a high-Q cavity, as shown in Fig. \ref{fig1}a1. Reflection or backscattering from the high-Q cavity locks the lasing frequency to the stable microcomb state region, enabling mode-locked microcombs under both anomalous\cite{shen2020integrated, briles2021hybrid, raja2019electrically} and normal\cite{jin2021hertz, lihachev2022platicon, gil2025high} dispersion in turnkey operations. However, this scheme still relies on CW pumping, where the lasing mode is only built at the overlapping of the laser cavity filtering and the high-Q cavity filtering. Other modes are stimulated through the Kerr oscillation excited by the pump laser. The separation of laser generation and Kerr oscillation constrains the overall conversion efficiency. In addition, an independent laser cavity is required, which adds complexity to the fabrication process, especially for heterogeneous integration\cite{xiang2021laser}. 

Beyond self-injection locking, hybrid cavities where a microcavity is embedded in a gain loop\cite{bao2020turing, nie2022dissipative, rowley2022self} have been proposed for self-sustained microcomb operation. One example is shown in Fig. \ref{fig1}a2. The high reflection facet of the gain chip and the reflection from the high quality microresonator form a laser cavity. Dual rings are arranged between these two reflectors, selecting one longitude mode for lasing \cite{stern2018battery}. This method eliminates the requirement of an independent laser cavity. However, the dual rings’ filtering limits the microcomb operation to single wavelength lasing pump concept, with limited conversion efficiency. Another hybrid cavity structure gets rid of the dual rings. The lasing field is directly built and resonates in the high-Q cavity, stimulating the optical nonlinearity for microcomb generation\cite{wang2017repetition}. By eliminating the need for a CW pump, these systems can theoretically realize 100\% utilization of optical power contributing to comb generation, with reports of up to 75\% mode efficiency experimentally\cite{bao2019laser}. While such implementations are often bulky, relying on fiber-based components such as isolators and erbium-doped fiber amplifiers\cite{rowley2022self, nie2024turnkey}, thereby limiting their potential for on-chip integration. To approach this goal, researchers have very recently demonstrated hybrid integrated microcomb lasers based on thin-film lithium niobate platform\cite{ling2024electrically}. However, achieving wideband microcomb generation in such systems still requires the injection of an external microwave signal provided by a source meter or an oscillation loop, which in turn demands additional auxiliary components, including high-speed photodiode, electrical amplifier, filter and phase shifter. Such an arrangement increases the system’s overall power consumption and remains far from being compatible with integrated implementation. To our knowledge, a fully integrated hybrid cavity for self-sustained microcomb lasing, featuring a simplified setup and operation, has yet to be demonstrated.

In this work, we demonstrate a fully integrated hybrid cavity for self-sustained microcomb operation that delivers high coherence, turnkey operation and long-term stability. The hybrid cavity consists of a microresonator with engineered intracavity reflection directly coupled to a reflective semiconductor optical amplifier (RSOA), as illustrated in Fig. \ref{fig1}a. Building on our previous work \cite{shen2024reliable}, we demonstrate that high quality factors of $4 \times 10^6$ and a suitable reflection ratio around 50\% can be achieved by introducing an intracavity coupler. Solely relying on the co-oscillation of the lasing process and Kerr nonlinearity, coherent microcomb states are observed both in simulation based on Ikeda map, and realized experimentally within a simple setup. We further evaluate the coherence of individual comb lines, revealing intrinsic linewidths below 1 kHz and integrated linewidths (at 1 ms) around 100 kHz, comparable to those achieved with self-injection-locked microcombs on the same platform \cite{shen2024reliable}. Moreover, with appropriate presetting, the coherent microcomb state can start via turnkey operation and maintained for over a day without active feedback. This approach offers a compact, simple, and power-efficient solution for generating highly coherent multi-wavelength sources, with promising applications in metrology, communications, sensing, computing, and beyond.

\begin{figure*}[ht]
\centering
\includegraphics[width = 18cm]{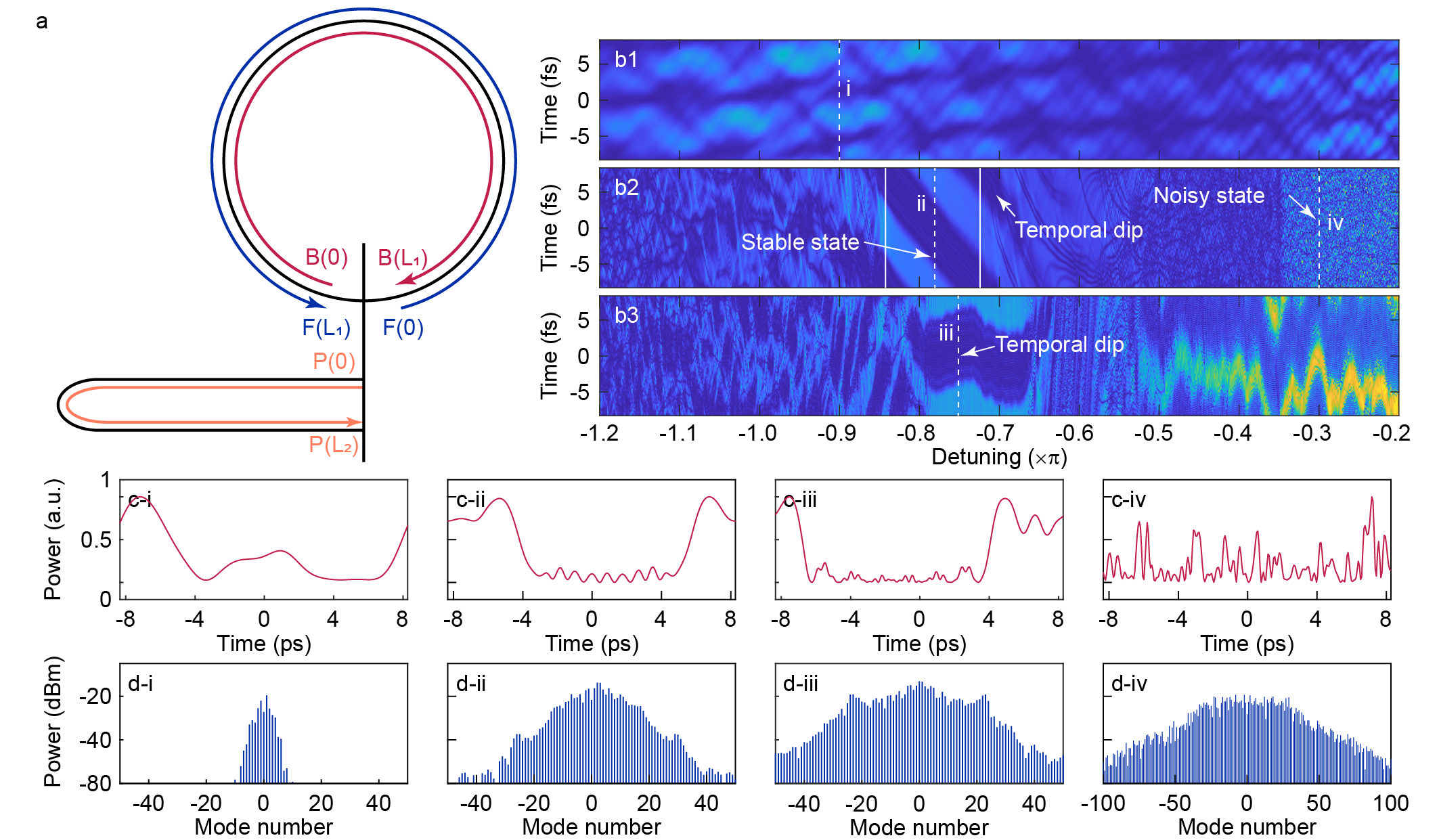}
\caption{\textbf{Microcomb generation in simulation. a,} The concept of the model. \textbf{b1-3,} The simulated temporal field evolution with the phase detuning under saturation powers of 0.02, 2.5 and 10 mW. \textbf{c,} The temporal fields at the marked detuning in Fig. 2b1-3. \textbf{d,} The optical spectra at the marked detuning in Fig. 2b1-3.
}
\label{fig2}
\end{figure*}

\begin{figure*}[ht]
\centering
\includegraphics[width = 18cm]{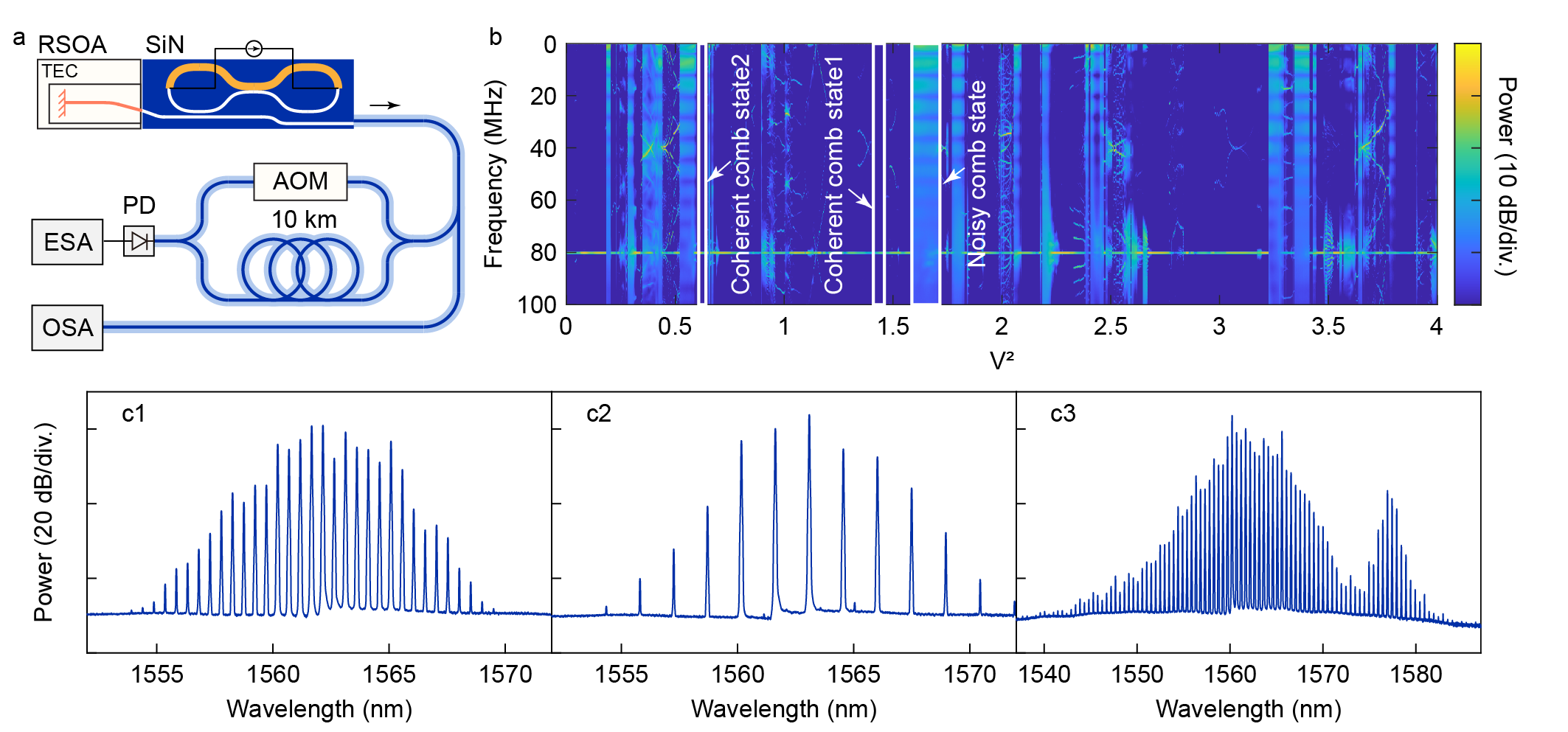}
\caption{\textbf{Microcomb generation in experiment. a,} Test link. \textbf{b,} The recorded beat spectra with the tuning of the applied voltage on the phase shifter. The optical spectrum of coherent comb state 1 (c1), 2 (c2) and noisy comb state (c3).
}
\label{fig3}
\end{figure*}

\begin{figure*}[ht]
\centering
\includegraphics[width = 18cm]{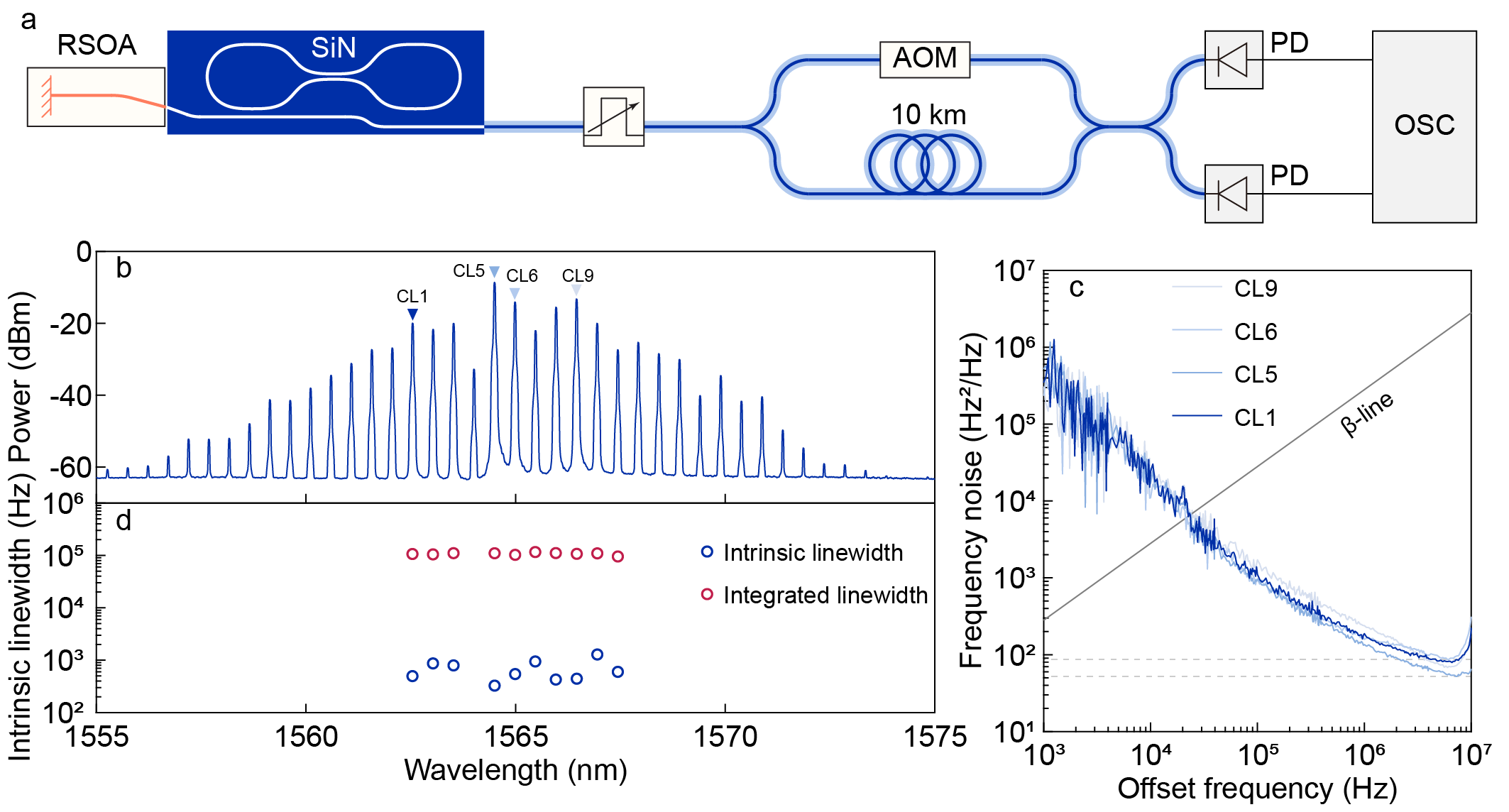}
\caption{\textbf{Coherence characterization. a,} Test link. \textbf{b,} The optical spectrum of a coherent microcomb state. \textbf{c,} The frequency noise profile of different comblines. \textbf{d,} The measured integrated linewidths (within 1 ms) and intrinsic linewidths of different comblines.
}
\label{fig4}
\end{figure*}

\begin{figure*}[ht]
\centering
\includegraphics[width = 18cm]{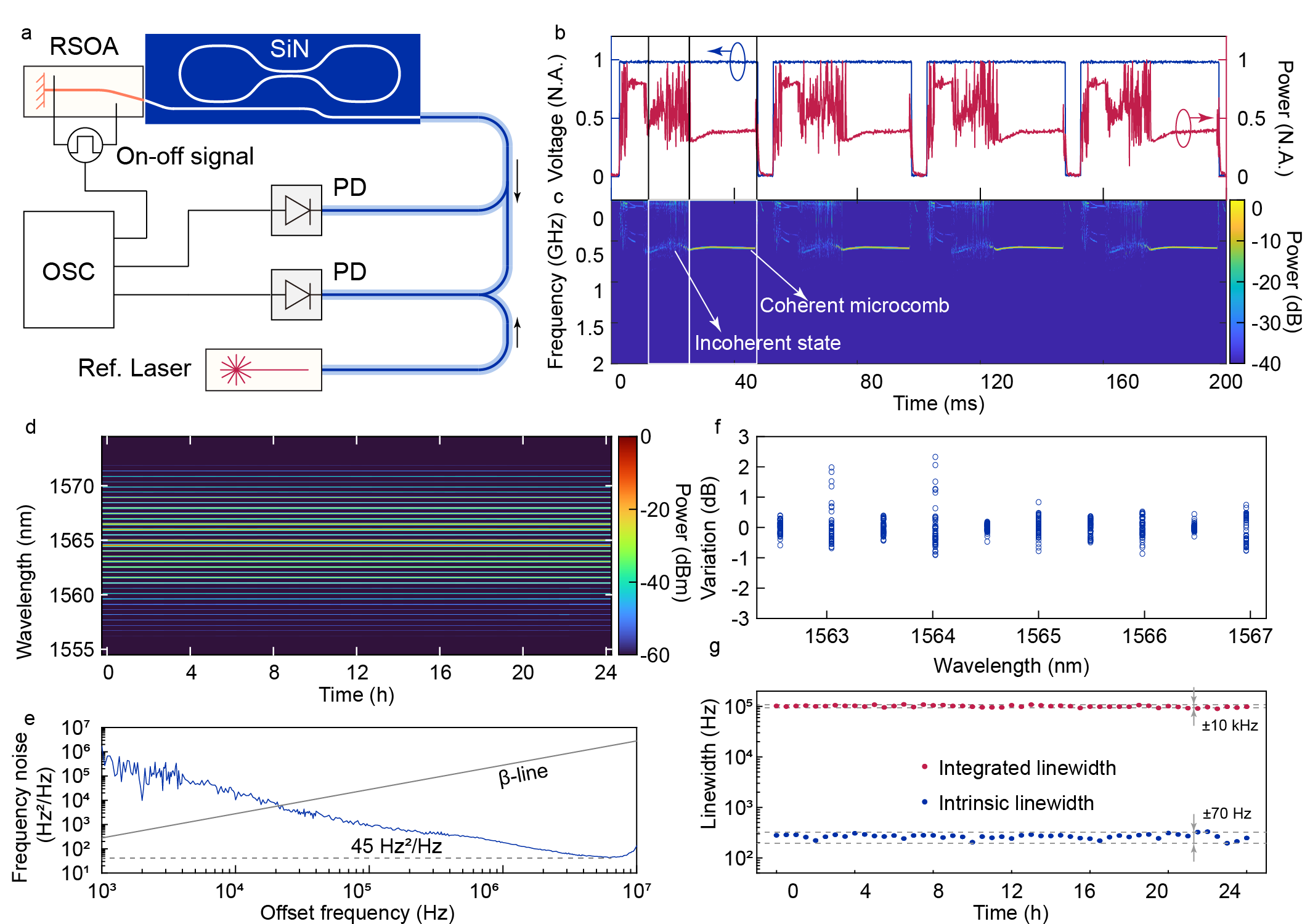}
\caption{\textbf{Turnkey operation and long-term stability. a,} Test link. \textbf{b,} The comb power variation with time. \textbf{c,} The spectrogram of the beating signal between the output of the hybrid cavity and the reference laser. \textbf{d,} The optical spectrum during a day. \textbf{e,} The frequency noise profile of one combline. \textbf{f,} The combline power variation during a day. \textbf{g,} The intrinsic linewidth and integrated linewidth (within 1 ms) variation during a day.
}
\label{fig5}
\end{figure*}

\vspace{3pt}
\noindent\textbf{Results} \\
\noindent\textbf{Device design and characterization} \\
\noindent Figure. \ref{fig1}a3 illustrates the operating principle of the proposed hybrid microcomb oscillator. The system consists of a reflective semiconductor optical amplifier (RSOA) and a SiN chip, within a \textless 2 mm$^2$ footprint (see Methods), as shown in Fig. \ref{fig1}b. The gain chip presents a flat gain spectrum centered at 1540 nm, from 1450 nm to 1600 nm (see Supplementary Information I). On the SiN chip, a high-quality-factor microring resonator, integrated with an intracavity directional coupler, serves dual functions: it acts as both an optical nonlinear cavity and a wavelength-selective reflector. Lasing oscillation can be constructed as the roundtrip gain is higher than loss in the main laser cavity, formed between the high-reflection facet of the RSOA and the reflection of the microring. With proper design\cite{shen2024reliable}, narrow reflection peaks are only formed around the resonant modes. This resonant feedback mechanism not only defines the lasing mode but also ensures that the intracavity field builds up near the microring’s resonance condition, where strong field enhancement and nonlinear optical effects can occur. Consequently, once lasing is initiated, the high circulating optical power within the microring can stimulate nonlinear processes, such as Kerr-induced four-wave mixing, leading to the generation of an optical frequency comb.

The microresonator plays a central role in enabling and shaping the performance of the hybrid oscillator. The chip is consisted of a 300 nm thick SiN on the isolator. To ensure a high quality, the waveguide width of the microresonator is 3 \textmu m, to reducing the sidewall scattering. At the center of the microresonator, the waveguide are bent to form an intracavity directional coupler, with the waveguide gap of 2 \textmu m. Such intracavity coupler can introduce coupling between the clockwise (CW) and counterclockwise (CCW) propagating modes, as illustrated in our previous work \cite{shen2024reliable}. In simulation, the reflection ratio induced by the coupler is $3\times10^{-5}$, leading to an intracavity CW-CCW coupling coefficient $\beta$ of 0.0055. The detailed structural parameters are provided in Supplementary note I. Figure. \ref{fig1}c and d shows the measured transmission and reflection spectra of the microresonator. The spectra reveal consistent extinction and reflection ratios across 1520-1600 nm. Two dips at 1545 nm and 1575 nm are observed in the reflection spectrum. These are attributed to inter-mode coupling between different transverse modes. Except for these two dips, the reflection ratios remains approximately constant around 50\%, ensuring effective wavelength-selective feedback across a broad spectrum. Figure. \ref{fig1}e shows the integrated dispersion profile $D_{int}(\mu)$, which is defined as $D_{int}(\mu)=\omega_\mu-\omega_0-D_1\cdot \mu=(D_2 \cdot \mu^2)⁄2+(D_3 \cdot \mu^3)⁄6+\cdots$, where $\mu$ is the index of the resonance, $\omega_\mu$ is angular frequency of the $\mu^{th}$ resonance, $D_k$ is the $k^{th}$-order dispersion coefficient. By fitting the measured data to this polynomial form, we extract the FSR as $D_1⁄2\pi$=59.60 GHz, and the second-order dispersion as $D_2⁄2\pi$=-3.20 MHz, indicating a strong normal dispersion. Figure. \ref{fig1}f shows one zoomed-in resonance near 1565 nm. A clear resonance splitting is observed, resulting from the CW–CCW mode coupling. The two split modes remain spectrally close, suggesting that the mode coupling rate is comparable to the total cavity loss rate. Such weak mode splitting is advantageous for achieving a stable state \cite{wang2025rhythmic}. The loaded quality factors are extracted by fitting the resonance peaks, as shown in Fig. \ref{fig1}g. The loaded quality factors around 1565 nm are about $4\times10^6$. Simultaneously achieving a broadband high reflection ratio and high quality factors ensures efficient coupling of the RSOA emission into the desired feedback mode for lasing and subsequent comb generation.

\vspace{3pt}
\noindent\textbf{Microcomb generation in the hybrid cavity}\\
\noindent In this section, we describe the microcomb generation process both in simulation and experiment. In simulation, we model the oscillation process using the Ikeda map \cite{hansson2016dynamics}, as illustrated in Fig. \ref{fig2}a. The model consists of two main components: a closed loop of the microring and an open loop coupled to the microring. For simplicity, we assume even-distributed coupling between the CW and CCW fields in the microring \cite{kippenberg2002modal}, and identical roundtrip lengths $L_1$ and $L_2$ of these two loops. The corresponding equations are given in the Methods. As illustrated in the preceding section, the co-oscillation of the lasing and the Kerr nonlinear process is essential for achieving a stable microcomb state. In principle, the gain factor in the open loop should be strong enough to compensate for the roundtrip loss. In this model, a saturable gain model \cite{bao2019laser} is employed, considering a maximum gain of $G_{max}$ and a saturation power of $P_{sat}$, where the gain is the half of the $G_{max}$. As shown in Fig. \ref{fig2}b, by sweeping the roundtrip phase difference between the two loops, equivalent to sweeping the detuning between the laser and microring cavities, we observe various states. As the $G_{max}$ is fixed at 500, corresponding to the 27 dB gain of the RSOA employed, we only get a lasing state as the $P_{sat}$ = 0.02 mW. This is due to insufficient intracavity power for Kerr nonlinear oscillation. Considering the normal dispersion, the low noise states should express as stable temporal dips in time domain, which are referred as dark pulse\cite{xue2015mode} or platicon\cite{lihachev2022platicon} microcomb states. Such state is observed as the $P_{sat}$ increases to 2.5 mW, marked in Fig. \ref{fig2}b2. The corresponding time-domain shape and the optical spectrum are also given in Fig. \ref{fig2}c-ii and d-ii. With a higher $P_{sat}$ = 10 mW, we can get a temporal dip state with a wider optical spectrum as shown in Fig. \ref{fig2}d-iii. While the generated temporal dip is unstable under a long-term evolution as shown in Supplementary Note III. Apart from the temporal dip state, noisy states are also observed in Fig. \ref{fig2}b with a wider optical spectrum. The corresponding temporal shape and the optical spectrum are given in Fig. \ref{fig2}c-iv and d-iv. According to the simulation results, the gain process plays a critical role in achieving a stable microcomb state. A low saturation power results in insufficient intracavity power for nonlinear oscillation, while a high saturation power makes the intracavity microcomb state unstable. It is worth noting that, the CW-CCW coupling rate $\beta$ also plays an important role as given in Supplementary Note III. 

In the experimental setup, we also achieve the microcomb state by varying the detuning between the laser cavity and the microring. To ensure suitable intrcavity power, the current applied to the RSOA is set to 200 mA. The detuning is controlled by adjusting the voltage applied to the thermal phase shifter on the microring, as shown in Fig. \ref{fig3}a. Optical spectra for different voltage settings are recorded using an optical spectrum analyzer (OSA). Additionally, the output of the hybrid cavity is sent through a delayed self-heterodyne setup and recorded by a OSC (high-speed oscilloscope). The spectra obtained by Fourier transforming the recorded signal are used to valuate the coherence of the generated states, as shown in Fig. \ref{fig3}b. Because the free-spectral range (FSR) of the microresonator is approximately nine times of the main laser cavity (6.6 GHz), the beatnote spectrum undergoes multiple transitions between noisy and clean single-tone state, which generally corresponds to the transition to high coherence state. Figure \ref{fig3}c presents optical spectra for three different states observed during the tuning process, marked in Fig. \ref{fig3}b. In the microcomb state shown in Fig. \ref{fig3}c1, the comb tooth spacing closely matches the free spectral range (FSR) of the microring. The self-heterodyne beat note shows a single tone at 80 MHz, correspondinig to the shifting frequency of the AOM (acousto-optic modulator), indicating a stable state. In contrast, the state in Fig. \ref{fig3}c3 exhibits a comb tooth spacing close to the FSR, but the self-heterodyne beat note displays a noisy spectrum, suggesting an incoherent state. Such noisy state expresses a wider optical spectrum compared with the coherent state, similarly to the simulation result. Additionally, we observe a microcomb state with a tooth spacing near three times of the FSR, which also produces a distinct single tone in the self-heterodyne beat note. This state may be attributed to localized dispersion variations arising from the coupling between different modes.

In the present setup, the microcomb exhibits a spectrum span of approximately 15 nm. The bandwidth is primarily limited by the normal dispersion condition. Compared with the bright soliton, operating under anomalous dispersion, the dark-pulse microcomb state is more fabrication-friendly and therefore more feasible for integrated implementations. Taking the SiN platform as an example, achieving anomalous dispersion near 1550 nm typically requires a film thickness of around 800 nm. However, depositing high-quality SiN layers thicker than 400 nm is challenging due to stress-induced cracking during fabrication. For this reason, we employ a 300-nm SiN platform to demonstrate the integrated hybrid cavity. Dark-pulse microcombs have been widely demonstrated for low-noise microwave generation \cite{sun2025chip}, high-capacity optical communications \cite{zhang2024high}, and optical processing \cite{shu2022microcomb}, with performance comparable to bright soliton microcombs.  We note that a broader spectral span is expected when operating under weaker normal-dispersion conditions, as discussed in Supplementary note IV.

\vspace{3pt}
\noindent\textbf{Coherence characterization}\\
\noindent
To quantitatively evaluate the coherence of the generated optical frequency comb, we characterized the frequency noise of individual comb teeth with a delayed self-heterodyne detection technique \cite{yuan2022correlated}. In this setup, the self-heterodyne signal is directed to two independent PDs (photodetectors). This signal is sampled using a OSC and subsequently processed to extract the frequency noise power spectral density (PSD). The coherent comb is generated under the same conditions as those shown in Fig. \ref{fig3}, except that the applied voltage on the phase shifter was removed to eliminate potential fluctuations. The optical spectrum is given in Fig. \ref{fig4}b. Frequency noise curves for four comb lines marked in Fig. \ref{fig4}b  are plotted in Figure \ref{fig4}c. Across all measured comb lines, the frequency noise PSD reaches a minimum at a frequency offset of approximately 8 MHz. This minimum defines the white frequency noise floor \cite{tran2019tutorial}, denoted as $S_0$, and the corresponding intrinsic linewidth can be estimated as $2 \pi S_0$. In addition to intrinsic linewidths, we also assess the 1-ms integrated linewidths using the $\beta$-separation line method \cite{di2010simple}, which offers a reliable estimate of phase noise over a given time scale. The results are summarized in Fig.\ref{fig4}d where both the intrinsic and integrated linewidths for selected comb teeth are presented, except for one comb line where the power is too weak for PDs. Among the measured lines, the comb tooth centered at 1564.4 nm exhibits the lowest intrinsic frequency noise, while surrounding comb teeth show slightly elevated noise levels. Nevertheless, all intrinsic linewidths remain below 1 kHz, demonstrating excellent coherence across the comb. The integrated linewidths are found to be relatively uniform across different wavelengths, approximately on the order of 100 kHz, further confirming consistent spectral coherence throughout the comb. Such high coherence performance is attributed to the high quality factor of the SiN microresonator. In laser cavities, a longer photon lifetime directly leads to higher coherence. In pure semiconductor laser cavities, the photon lifetime is typically limited to picosecond-level due to the intrinsic loss of III–V materials. In contrast, the photon lifetime of the SiN cavity in our design increases to $\sim$ 3 ns, enabling significantly reduced phase noise. This strategy has been widely adopted in integrated external-cavity lasers \cite{cuyvers2021low, dong2025sub}. Compared to self-injection-locked microcombs realized on the same platform \cite{shen2024reliable}, another well-demonstrated integrated microcomb lasing, the noise characteristics of the current device show comparable performance, validating the efficacy of our simplified integrated design.

\vspace{3pt}
\noindent\textbf{Turnkey operation and long-term stability}\\
\noindent In addition to the generation method based on tuning, such high coherence microcomb state can also be reliably achieved via a turnkey operation, provided that the system parameters, such as pump current, cavity feedback, and coupling conditions, are appropriately set. This greatly simplifies practical use, especially in scenarios requiring repeatable and stable comb generation without dynamic adjustment. As illustrated in Fig. \ref{fig5}a, a periodic on-off modulation signal is applied to the reflective semiconductor optical amplifier (RSOA) using an arbitrary function generator. This electrical signal controls the injection current to the RSOA, thereby switching the hybrid cavity system between inactive and active states. To monitor the evolution of the optical output, the light emitted from the hybrid cavity is directed into a photodetector, and the temporal variation of output power is recorded. To further evaluate the coherence of the generated state, the output is heterodyned with a reference continuous-wave (CW) laser, and the resulting beat note is analyzed. The dynamic formation of the microcomb is captured in Fig. \ref{fig5}b, which shows the photodetector voltage as a function of time. As the RSOA current ramps up, the output power increases and reaches a steady-state level after approximately 20 ms. Simultaneously, the spectrogram of the beating signal reveals a transition from a noisy, broadband beat to a well-defined single-tone frequency component centered around 0.5 GHz, as shown in Fig. \ref{fig5}c, indicating the transition from an incoherent state to a stable, coherent microcomb regime. The final optical spectrum of the generated state matches that shown in Fig. \ref{fig4}b, confirming the reproducibility of the target comb state.

Beyond the ease of generation, the long-term stability of the microcomb is also experimentally validated. During the test, the microcomb lasing module is placed on a temperature controller that maintained temperature fluctuations below 5 mK, ensuring a highly stable environment. Aside from the temperature controller, no other active feedback loops are employed. Figure. \ref{fig5}d displays the recorded optical spectra at 30-minute intervals over a continuous 24-hour period. The spectral envelope and comb line positions remain essentially unchanged, indicating excellent spectral stability. To quantify temporal power fluctuations, Fig. \ref{fig5}f shows the measured power variations of the central ten comb lines. The most intense line at 1564.4 nm shows a variation of only 0.47 dB, while a weaker adjacent line at 1564.0 nm experiences a largest variation of approximately 3.23 dB. These fluctuations are primarily attributed to environment temperature changes throughout the day. To mitigate such environmental sensitivity, hermetic vacuum packaging can be employed to isolate the device from thermal and atmospheric perturbations, further enhancing operational stability for real-world applications. In addition to power stability, we assess the frequency noise of the generated comb lines. Figure \ref{fig5}e shows the frequency noise spectrum of the 1564.4 nm line. By applying the standard $\beta$-separation line technique, both the intrinsic linewidth and the integrated linewidth over a 1 ms observation window are extracted. As shown in Fig. \ref{fig5}g, the intrinsic linewidth fluctuates around 200 Hz with a deviation of ±70 Hz, while the integrated linewidth remains centered near 100 kHz with a variation of ±10 kHz.

\vspace{3pt}
\noindent
\textbf{Conclusion and discussion} \\
In this work, we demonstrate a simplified, fully integrated hybrid cavity for self-sustained microcomb operation. The system features a high-Q microresonator with engineered intracavity reflection, directly coupled to a reflective semiconductor optical amplifier (RSOA). The generated microcomb exhibits high coherence, comparable to that of a self-injection locked microcomb on the same platform. Furthermore, due to the hybrid oscillation mechanism, the coherent microcomb is self-starting, enabling repeatable, turnkey operation.

In the main text, we focus on simulation results assuming equal roundtrip lengths for the cavities. However, as a coupled cavity system, variations in the roundtrip delay between the two cavities may affect the system's robustness \cite{ling2024electrically, rowley2022self}. This can be modeled by introducing first-order dispersion mismatch into the coupled equations, as detailed in Supplementary Note IV. With these modified equations, we track the evolution of a stable comb state as the roundtrip delay differences vary. We find that a stable microcomb state can persist with cavity delay variations of up to $\pm 0.5\%$, a mismatch that can be effectively controlled through the fabrication process, as demonstrated by 0.09\% variations reported in \cite{qiu2025large}.

In addition to the cavity length mismatch, several theoretical and practical questions remain to be addressed. For instance, the results presented so far are based on normal dispersion conditions. However, this configuration could potentially be extended to cavities with anomalous dispersion. The effect of dispersion—both normal and anomalous—and the possible influence of optical nonlinearities within the laser cavity require further investigation. These studies will provide a deeper understanding of the system's performance and help optimize hybrid cavity designs for a broader range of applications.

\vspace{6pt}
\noindent \textbf{Methods}\\
\begin{footnotesize}
\noindent \textbf{Ikeda map for the hybrid cavity.} 
As illustrated in Fig. 2a, the evolution in the hybrid cavity can be modeled by considering the dynamics within two loops and the coupling between them. In the microresonator, the field evolution can be described by the nonlinear Schrodinger equations, which accounts for the intracavity coupling between the clockwise (CW) and counterclockwise (CCW) fields \cite{kondratiev2020modulational}:
\begin{equation}
\frac{\partial F}{\partial z} = -\frac{\alpha}{2} F - i \frac{\beta_{21}}{2} \frac{\partial^2 F}{\partial \tau^2} + i \gamma \left( |F|^2 + 2P_B \right) F + i \beta \cdot B(-\tau)
\end{equation}
\begin{equation}
\frac{\partial B}{\partial z} = -\frac{\alpha}{2} B - i \frac{\beta_{22}}{2} \frac{\partial^2 B}{\partial \tau^2} + i \gamma \left( |B|^2 + 2P_F \right) B + i \beta \cdot F(-\tau)
\end{equation}
where $F$ and $B$ are the CCW and CW fields respectively, $\alpha$ represents the roundtrip power loss factor, $\beta_{21}$ is the second order dispersion coefficient in cavity 1, $\tau$ is the fast evolution time, $z$ is the normalized propagation length, $\gamma$ is the kerr nonlinear coefficient, $\beta$ is the coupling coefficient between CCW and CW field, $P_F$ and $P_B$ represents the average powers of CCW and CW field.

For the open loop containing the gain medium, the evolution can be modeled as:

\begin{equation}
\frac{\partial P}{\partial z} = \left( -\frac{\alpha'}{2} + g + i \delta \right) P - i \Delta \frac{\partial P}{\partial \tau} - i \frac{\beta_{22}}{2} \frac{\partial^2 P}{\partial \tau^2}
\end{equation}
where $P$ is the field in the open loop, $\alpha’$ is the roundtrip power loss factor in the open loop, $g$ is the gain factor, $\delta$ is the roundtrip phase mismatch, $\beta_{22}$ is the second order dispersion coefficient. In the manuscript, we employ a saturation gain model to simulate the lasing process. The gain factor g is given as:
\begin{equation}
g=\frac{g_{\max }}{1+\frac{|A|^2}{P_{\text {sat }}}}
\end{equation}
where $g_{max}=\sqrt{G_{max}}$ is the field gain factor and $P_{sat}$ is the saturation power. It is worth noting that nonlinear effects in the open loop are neglected for this model. At the coupling point between the resonator and the open loop, the field are updated as follows:

\begin{equation}
F_{m+1}(0, \tau) = i \sqrt{\theta} P_m(L_2, \tau) + \sqrt{1 - \theta} F_m(L_1, \tau)
\end{equation}
\begin{equation}
B_{m+1}(0, \tau) = \sqrt{1 - \theta} B_m(L_1, \tau)
\end{equation}
\begin{equation}
P_{m+1}(0, \tau) = i \sqrt{\theta} B_m(L_1, -\tau)
\end{equation}
where $\theta$ is the coupling coefficient of the coupler, $m$ is the roundtrip number, $L_1$ and $L_2$ are the roundtrip length of the resonator and the open loop. The entire evolution process in the hybrid cavity is described by Eq. 1-6. The output field $A$ is given by:

\begin{equation}
A_{m+1} = \sqrt{1 - \theta} P_m(L_2, \tau) + i \sqrt{\theta} F_m(L_1, \tau)
\end{equation}
For simple expression, we define:
\begin{equation}
\beta = \beta ' \cdot \frac{\alpha + \theta}{2}
\end{equation}

\noindent \textbf{SiN chip structure.} 
The SiN chip is fabricated by CUMEC. The detailed design of the microresonator is given in Supplementary Note I. The thickness of the aluminum electrode on the microring is 0.6 \textmu m, 2 \textmu m above the SiN layer. The area within the orange box shown in Fig. \ref{fig1}b is 1.08 mm$^2$, including the edge coupler, the microresonator and pads.

\noindent \textbf{Setup for microcomb generation and characterization.} 
In the experiment, we use a commercial RSOA chip (Anritsu AE5T310BY10P) as the optical gain medium. The size of the RSOA chip is 1 mm \texttimes 0.4 mm. The RSOA is driven and controlled by a Thorlabs ITC4001 laser controller. For microcomb generation, the current of the RSOA is set to 200 mA with the voltage of 1.4 V, and the temperature is kept at 23.64 °C. The voltage applied to the phase shifter is varied from 0 to 2 V, with higher voltages potentially damaging the phase shifter. As discussed in Supplementary Note II, this voltage range is insufficient to ensure that the resonance sweeps through a full FSR of the resonator. Because the cavity length of the laser circuit is several times that of the microring, we can still observe several transitions from coherent to incoherent states.

For frequency noise measurement, two identical low-noise photodetectors are used. The AOM is driven by an 80 MHz microwave signal to shift the corresponding signal by 80 MHz. In the coherent state, the output of the PD should exhibit a sine wave signal at 80 MHz. Using the method proposed in \cite{jin2021hertz}, we can derive the frequency noise profile of the injected sources.

\end{footnotesize}

\begin{footnotesize}
\vspace{3pt}
\noindent\textbf{Data availability}\\
The data that supports the plots within this paper and other findings of this study are available from the corresponding authors upon reasonable request. 

\vspace{3pt}
\noindent\textbf{Code availability}\\
The codes that support the findings of this study are available from the corresponding authors upon reasonable request.

\vspace{3pt}
\noindent \textbf{Acknowledgments}\\
This work was supported by  National Key R\&D Program of China (2022YFB2802400), National Natural Science Foundation of China under Grant (62235002, 62327811, 62322501, 12204021), China National Postdoctoral Program for Innovative Talents (BX20240014) and High-performance Computing Platform of Peking University. The authors thank Zhangfeng Ge in Peking University Yangtze Delta Institute of Optoelectronics for providing equipments.

\vspace{3pt}
\noindent \textbf{Author contributions}\\
The idea was conceived by B.S. The device simulation was established by B.S., H.C. and J.H. The simulation model was established by B.S. with assistance by J.H. The experimental setup was performed by B.S., H.C. and J.H. The experiment was performed by B.S, H.C. and J.H. with assistance by Y.W., X.Z., H.W., Z.T. and R.C. The results were analyzed by B.S., H.C. and J.H. All authors participated in writing the manuscript. The project was under the supervision of H.S. and X.W.

\vspace{3pt}
\noindent \textbf{Competing financial interests} \\
The authors declare no competing interests.

\vspace{3pt}
\noindent \textbf{Additional information} \\
Supplementary information is available in the online version of the paper. Reprints and permissions information is available online. Correspondence and requests for materials should be addressed to H.W., and X.W.

\vspace{6pt}
\noindent \textbf{Corresponding authors}\\
Correspondence to Haowen Shu and Xingjun Wang.
\end{footnotesize}

\vspace{20pt}

\bibliography{REF.bib}

\end{document}